\documentclass[11pt]{article}
\usepackage{color,graphicx}
\usepackage{amsmath}
\usepackage{amssymb}
\usepackage{amsfonts}
\newcommand{\mtt} {M_{t\bar t}}
\newcommand{\ttbar}{t {\bar t}}

\begin{document}

\title{Keynote: Some remarks on top\footnote{Talk given at the
  3rd International Workshop on Top Quark Physics, 
May 31 - June 4, 2010, Bruges, Belgium}}
\author{Werner~Bernreuther \\
 Institut f. Theoretische Physik, RWTH Aachen
University -  52056 Aachen, Germany}

\date{}

\maketitle

\begin{abstract}
A few key issues of present and future explorations of the 
   physics of top quarks at the Tevatron and LHC are discussed.
\end{abstract}

\section{Where do we stand?}
It is no exaggeration to state that
  significant  progress has been made in the exploration   of
  top quarks  
since the {\it Top2008} workshop at  La Biodola/Elba.
 During the last two years, an
   impressive number of new results have been obtained by the 
   CDF  and D$\emptyset$ experiments at the Tevatron.
 Let me mention a few highlights.
 The measurements of $\sigma_{\ttbar}$ in the main 
  channels have been
   improved. The experimental uncertainty now reaches $\sim 6.5\%$
  \cite{Fdeliot}. 
    The measurements of the top mass have become more precise than
       anticipated several years ago.
        The  uncertainty of the  CDF  and D$\emptyset$
          average as of 2009  is $0.75 \%$ \cite{:2009ec,Obrandt} --
        it is  the most precisely
   known quark mass  (and the value of $m_t^{exp}$ ``converges''). 
         Single top-quark production at the Tevatron made it from
        evidence to  being observed 
        \cite{Abazov:2009ii,Aaltonen:2009jj,Jluck,Aheinson}. 
          The experimental information about 
        top-quark decays -- or to put it differently, the information
         about those modes into which the top quarks produced
       so far have not decayed -- has been
         refined:   $t\to b W$ is still the only decay mode observed.
     From the measurements of  the single top cross section and of
     the branching ratios $B(t\to b W(h_W=0,\mp))$ one can conclude that
      the strength and structure of   the $tWb$ vertex is known now
         to a
      precision of about $10 - 20 \%$  \cite{datta,Aharel}.  Quite 
              recently,  
     the experimental knowledge about the total width $\Gamma_t$
     of the top quark has been improved \cite{datta,Aharel}.
      As far as the analysis of the $t \bar t$ events
         at the Tevatron is concerned, a number of distributions
         have been measured, including the top-quark charge asymmetry 
        \cite{:2007qb,Aaltonen:2008hc,CDFpublic1,datta}
         and $t \bar t$ spin correlations 
               \cite{D0public1,CDFpublic2,CDFpublic3,Head}. 
             The $\ttbar$ invariant-mass
         spectrum has been explored up to 
          about  $\mtt \sim 1$ TeV 
       in the (so far negative) search  for heavy, electrically
           neutral resonances
                  that (strongly) couple to top quarks \cite{sinervo}.  Last but not
                  least, it should be appreciated that quite a number
                  of methods that were developed to analyze $\ttbar$
                  and single top events \cite{Canelli} (will) 
                serve as templates in
                  the search for other heavy (colored) particles. 
          
Also theorists have not been idle in the past two years. There has
been an ever increasing number of papers  on top-quark phenomenology,
both within the standard model (SM) and beyond. 
Already at the  {\it Top2008} workshop, theory updates (NLO QCD plus
  NLL threshold resummations) of
 the $\ttbar$ cross sections for the Tevatron and the LHC were
 presented \cite{Moch:2008qy,Cacciari:2008zb,Kidonakis:2008mu}. 
    More recently, threshold resummations were
  extended to NNLL order for the partonic cross 
sections \cite{Czakon:2009zw,Beneke:2009ye}
   and for the $\ttbar$ invariant-mass distribution \cite{Ahrens:2010zv}.
 The $\ttbar$ cross section was computed in terms of the running 
 ${\overline{\rm MS}}$ top mass and its value was extracted from the measured
  Tevatron cross section \cite{Langenfeld:2009wd}.
 These and related
    issues, including building blocks obtained so far
  for the
   computation of $\sigma_{\ttbar}$ to order $\alpha_s^4$,  will 
 be discussed at this workshop by \cite{SMoch}.
  The formation of
   a smeared $\ttbar$ resonance peak in $gg\to \ttbar$ at threshold
    and the resulting distortion of the $\mtt$
   distribution at $\mtt\approx 2 m_t$ was analyzed for
        top-quark pair production at the LHC \cite{Hagiwara:2008df,Yokoya}. It is an
   interesting effect; yet it seems not possible with present
   state-of-the-art methods to experimentally resolve it.
  A number of refined SM predictions were made for distributions that
   can be measured in $\ttbar$ events, including the top-quark
   charge asymmetry in threshold resummed (NLL \cite{Almeida:2008ug} and NNLL \cite{Ahrens:2010zv})
          QCD perturbation theory, charge asymmetries in dileptonic
          final states \cite{Bernreuther:2010ny},  and
     final-state angular correlations induced by $\ttbar$ spin
    correlations at next-to-leading order in the strong and weak gauge 
    couplings \cite{Bernreuther:2010ny}. Recently, an interesting suggestion 
      was made how  spin correlations for
    low-mass dileptonic $\ttbar$ events can be measured at the
    LHC, once enough events will have been collected
      \cite{Mahlon:2010gw,Mahlon}.
    Concerning the general-purpose NLO QCD Monte-Carlo generators
     MC@NLO, MCFM, and POWHEG, a number of new features  were added 
         to these programs in   recent years concerning 
      reactions that involve top quarks. 
   This will be discussed by \cite{PNason,Cwhite}.
     Results based on computer programs specific to hadronic
     top-quark pair production {\it and} decay at NLO QCD including spin
     correlations were  reported \cite{Melnikov:2009dn,Bernreuther:2010ny}. (The investigation in
        \cite{Bernreuther:2010ny} includes also weak-interaction corrections.)
     The $W$-boson helicity fractions in $t\to bW$ decay
      were computed to NNLO QCD \cite{Czarnecki:2010gb}. 
          Concerning hadronic single top
     production the dominant $t$-channel production cross section and 
      distributions were 
     determined to  NLO QCD in the 4-flavor scheme and compared with
     corresponding results in the 5-flavor scheme \cite{Campbell:2009gj,Rfrederix}. This is important
     for assessing the uncertainties of the SM predictions. 
      Phenomenological studies include the development of algorithms
      to analyze high-$p_T$ top-quark events (``boosted tops'') 
    \cite{Kaplan:2008ie,Almeida:2008tp,chabert} and
       studies for  determining  possible anomalous
      couplings in the $tWb$ vertex from data on single top
          and $\ttbar$ production \cite{AguilarSaavedra:2010nx}.

 There have been many phenomenological investigations
     on effects beyond the standard model (BSM) in top-quark
     production and decay. For instance, the measurements of the
     top-quark charge asymmetry at the Tevatron by  D$\emptyset$
       and especially by CDF, which do not quite match the
       QCD predictions, have induced a plethora of papers on possible
         new physics contributions to this observable. This will be 
     discussed by \cite{Rodrigo}. An issue that has recently been revived is the
      possible existence of a fourth sequential heavy quark
      generation \cite{Hou}, or of more exotic, e.g. vectorlike
       quarks. The existence of such quarks
        would have an impact on top-quark physics, too, see below.
        A central theme is and will be the use of top quarks as 
         a probe of the hitherto unknown mechanism of electroweak
         gauge symmetry breaking. In a more general  context many 
          studies were made on heavy BSM resonances that strongly
         couple to top quarks. A few comments  on this topic  will be
         made below.  
                 Furthermore, the issue 
           of how dileptonic and semileptonic $\ttbar$ events
          can be used to search for non-standard CP
         violation has been taken up again \cite{Valencia}, in view of 
           the forthcoming LHC data samples.

At this workshop we are mainly interested in the physics of ``top as a
 signal''. But top-quark production and decay constitutes also an
 important background to a number of new physics
  searches, including searches for (non-SM)  Higgs boson(s) and
     SUSY particles. In order to understand and 
 control this background,  precise SM predictions for the
 respective top quark reactions are required.  For the
    most important background reactions that involve $\ttbar$, 
     predictions of the cross
     sections and of distributions are now available 
      at NLO QCD, namely for $\ttbar + {\rm jet}$ \cite{Dittmaier:2008uj,Melnikov:2010iu},
      $\ttbar + {b \bar b}$ 
\cite{Bredenstein:2009aj,Bredenstein:2010rs,Bevilacqua:2009zn}, and  $\ttbar + 2 \, {\rm jets}$
    \cite{Bevilacqua:2010ve},    see the talk by \cite{Mworek}.

   To sum up the present state of the art of top physics:   
  The results of the CDF and  D$\emptyset$
      experiments imply that the top quark
     behaves pretty much standard. (The measurements of the
      charge/forward-backward asymmetry may point to an exception.)
      On the theory side, the main $ttX$ and single top processes 
        were computed to NLO in the
  SM gauge couplings, and many options for BSM effects have been
     studied. 
   
 As to  present and future top-physics issues at the 
 Tevatron and the LHC: Needless to say, the analysis of increased data
   samples will further sharpen the profile of this quark. Besides
    further
     determinations of
        its already  very precisely known mass, direct 
  measurements respectively extractions of its charge and spin 
        will be possible. We expect to eventually 
           obtain  very detailed knowledge about 
   the top-quark decay modes and perhaps also of its width --
  although no method is known to  
     {\it directly} measure the  top width $\Gamma_t$
  at a hadron collider with a precision of, say, $\lesssim 40 \%$. 
  (From what is presently  known about the top quark, one  can conclude
      that its width  $\Gamma_t$
       cannot differ, if at all,  more than about
         $40\%$ from $\Gamma_t^{SM} \simeq 1.4$ GeV.) 
  There will  be more detailed measurements and theoretical
       investigations   of the cross sections and distributions 
         for the main reaction channels. From a more general point
     of view we expect that top-quark physics will significantly
      contribute to gaining insights into
           two grand questions of particle physics, namely the 
         flavor problem and the issue of electroweak gauge
         symmetry breaking. Flavor-physics aspects, already pursued at
         the Tevatron,  include
        the search for new decay modes, e.g. $t\to {\tilde t} {\tilde \chi}^0$,
       $t\to H^+ b$, for FCNC decays $t \to c,u$
   and for  detectable FCNC in top production,  
   $p {\bar p}, pp \to t{\bar c} \, X, \,  t{\bar u} \,X$, and 
         further searches  for an  
  additional sequential quark generation or exotic
 heavy quarks which may mix with the top quark. We expect
   that the LHC  experiments  will eventually be able to explore
    also top's capability  to
      probe the  mechanism of electroweak    
    gauge-symmetry breaking (EWSB) by its coupling 
    to the SM Higgs boson (if it exists) or possibly by its
        couplings to other spin-zero resonances from the EWSB
         sector.

   These themes will be discussed in detail in the forthcoming
   talks at this workshop. Here I shall restrict 
   myself to some comments on a few selected topics. 
    
\section{Remarks on selected topics}
\subsection{Top mass}
 CDF and  D$\emptyset$ have
 precisely determined the top mass
 by exploiting  the $\ttbar$ event kinematics, 
using Born matrix element, template, and ideogram methods.
  The CDF  and D$\emptyset$ average as
  of 2009 is $m_t^{exp} \, = \, 173.1 \, \pm \, 1.3$ GeV
        \cite{:2009ec,Obrandt}. This
  average has  an error of 0.75 \% -- but which mass 
 is measured, i.e., how does
          $m_t^{exp}$  relate 
   to a (well-defined) quark  mass parameter used in quantum field theory?
     We had this discussion already at the {\it Top2008}
   workshop \cite{Hoang:2008zz}. Obviously, $m_t^{exp}$ is the mass parameter
  in the Monte Carlo programs with which the experiments 
   form their templates, etc. that are fitted to their data.
    It is reasonable to identify
  $m_t^{exp}$ with the pole or on-shell mass 
  $m_t^{on}$ --  
but this cannot be completely correct, because the 
   top quark is a colored resonance, 
      while the data involve color-singlet final states.   
The determination of $m_t^{exp}$  is hard to map onto a 
 higher order QCD calculation.  

A well-known uncertainty which is
 involved in the  top-mass determinations from the 
   peak of the top invariant mass distribution and from fits to 
    perturbative (Born) matrix elements of the partonic
    reactions $q{\bar q}, gg \to {\ttbar} \to f$ are color 
 reconnection effects, i.e. the color exchange
between the $t$, $\bar t$ decay products (specifically
   $b$ and $\bar b$)  and the proton remnants. This is
   a non-perturbative QCD effect, which at present can be (and is) taken
   into account by heuristic Monte Carlo estimates, which yield
    an uncertainty $\delta m_t \sim 0.5$ GeV.
The {\it ab initio} calculation  of
    color reconnection effects in hadronic $\ttbar$ production 
   and  decay remains a challenge. 

A promising method, which does not suffer from this 
  problem, is the exploitation of the fact that
   the QCD cross section $\sigma_{\ttbar}$ varies with $m_t$
   as ${\Delta \sigma}/{\sigma} \simeq -5 {\Delta m_t}/{m_t}$,
   both for the Tevatron and  the LHC. Thus, the
   strategy is to compute  $\sigma_{\ttbar}$ in terms  
    of a short-distance mass,   e.g. 
   $m_t^{\overline{\rm MS}}$. (Unlike $m_t^{on}$ these mass 
   parameters are  well-defined.)
    The comparison of 
 $\sigma^{exp}_{t\bar t}$ and $\sigma^{th}_{t\bar t}$ then
  yields $m_t^{\overline{\rm MS}}$ \cite{Langenfeld:2009wd}.
     Comparing
    with the Tevatron cross section 
 the authors of \cite{Langenfeld:2009wd} extracted the
    running top-mass 
 ${\overline m_t}(\mu={\bar m_t}) = 160.0 \, \pm 3.3$ GeV.
   Converting this mass parameter to the on-shell mass
     yields $m_t^{on}=168.9 \, \pm \, 3.5$ GeV, which compares
    well with the CDF  and D$\emptyset$ average $m_t^{exp}$.
    It should be kept in mind, however, that SM production
     dynamics is assumed in the  computation of $\sigma_{\ttbar}$.
      In addition, the extraction of
    $\sigma^{exp}_{t\bar t}$  requires to correct for acceptance 
       cuts, which depend on the value of the top mass. If
     one aims at a more precise determination of  ${\overline m_t}$
    in the future, one should eventually compute in higher order QCD the
     $\ttbar$ cross sections for dileptonic and lepton + jets 
       final states with acceptance cuts.

 In the literature several other  kinematical methods
       for determining $m_t$ were proposed and 
    may be applied, especially in the high luminosity
       era of the LHC. A well-known suggestion  is
       the exploitation of $\ttbar \to b \, (\to J/\Psi \to \mu \mu) \,  + \ell \nu_\ell \; + \; {\rm jets}$, where 
 $m_t$ is correlated with the invariant  mass $M_{J/\Psi \ell}$
      \cite{Kharchilava:1999yj}.
  A similar variable is the invariant mass distribution
  $M_{\ell^+ j_b}$. (To leading order, 
   max $M_{\ell^+ j_b}^2 = m_t^2 - m_W^2$.) The sensitivity
   of this distribution to $m_t$ was recently studied in NLO
    QCD by \cite{Biswas:2010sa}. The decay length  of
        a $b$-hadron from top decay is also correlated
     with $m_t$ \cite{Hill:2005zy}. Moreover, as pointed out 
  in \cite{Frederix:2007gi}, the
    average of the $\ttbar$ invariant mass,  $\langle M_{\ttbar}\rangle$, and higher moments  are sensitive to the mass of the top quark.
     These methods have different experimental and theoretical 
     uncertainties (e.g. color reconnection, hadronization), which remain to be studied in detail   \cite{Corcella}.

\subsection{Strength and structure of $tWb$ vertex, new decay modes}
Here the issue is to test  the CKM
   universality of the charged weak quark current, i.e. the V-A
  law, in the  decay
  $t\to bW$. In the 3-generation SM, the respective branching fraction
     is $B(t\to b W)\simeq 99.9\%$. The D$\emptyset$ experiment has measured
      $B(t\to b W)=0.97^{+0.09}_{-0.08}$ \cite{Aharel}. The Lorentz structure
    of the $tWb$ vertex can be determined from the 
       $W$-boson helicity fractions $f_{0,\mp}$.  
   In the SM they are precisely known \cite{Do:2002ky,Czarnecki:2010gb}: 
   $f_0(h_W=0) \simeq 70 \%,$ $f_-(h_W=-1)\simeq 30 \%,$
  $f_+(h_W=+1)\simeq 0.1 \%$ $(f_0 + f_- + f_+ =1)$; the numbers 
 depend somewhat on the
     value of  $m_t$. (As an aside, it is worth pointing out that
     top-quark decay is the most copious source
          of longitudinally polarized $W$ bosons at the
          Tevatron and at the LHC.) These fractions
   are experimentally determined by measuring
   the $\cos\theta^*_\ell$, $M^2_{\ell b}$, $p_T^\ell$ distributions
    in semileptonic top-quark decay. The CDF and  D$\emptyset$ collaborations
    have performed 1- and 2-parameter fits to these distributions
    and obtained values for $f_{0,\mp}$ with errors of order $\delta f_{0,\mp} \sim 10 \%$. For details, see \cite{datta,Aharel}.\\
Lorentz covariance dictates that the  on-shell $t\to Wb$ amplitude
 depends on four form factors, two chirality-conserving ($f_L, f_R)$
    and two chirality-flipping ($g_L, g_R)$ ones, which may in general be complex.
In the SM with 3 quark generations we have, to Born approximation,
    $f_L = V_{tb}$ (modulo $g_W/\sqrt{2}$), i.e. $|f_L|=1$,  $f_R, g_L, g_R=0$. The
    Tevatron experiments
       have obtained bounds on these form factors from the measured
     $W$-boson helicity fractions  \cite{datta,Aharel}.
 There are strong indirect constraints on $f_R$ and $g_L$ 
 from decays $B\to
X_s \gamma$: $|f_R|, |g_L| \lesssim$ few $\times 10^{-3}$, but these
 bounds are not water-proof. 
Simulation studies  \cite{Hubaut:2005er,AguilarSaavedra:2007rs}
    for the LHC (14 TeV) with  10 fb$^{-1}$
  anticipate sensitivities
 $|\delta f_R| \gtrsim 0.06$,  
$|\delta g_L| \gtrsim 0.05$, $|\delta g_R| \gtrsim 0.03$.
  The strength of the dominant left-chiral form factor $f_L$ can be
  inferred from the measured single top cross section at the Tevatron.
  This yields  $f_L= 1.07 \pm 0.12$ (D$\emptyset$ \cite{Abazov:2009ii}) 
       and
    $f_L = 0.91 \pm 0.11 \pm 0.07$ (CDF \cite{Aaltonen:2009jj}). 
     At the LHC (14 TeV) it is
 expected to reach a sensitivity of $|\delta f_L| \sim 0.05$.
  At this point it should be emphasized that for a joint analysis
 of $\ttbar$ and single top production and decay data, a form factor
    decomposition of the $tWb$ vertex is no longer appropriate; 
    for a (relatively model-independent) gauge-invariant
    parameterization  of possible new physics effects one should use
     an effective Lagrangian which contains, in particular, the
      anomalous couplings $\delta f_L = f_L -1,$ $f_R,$ $g_L,$ and
        $g_R$ (see \cite{willenbrock}).

Computations of these form factors in a number of SM extensions
  with 3 quark generations (multi-Higgs and SUSY extensions, top-color
  assisted technicolor (TC2)) \cite{Bernreuther:2008us} yield that 
1-loop radiative corrections  induce non-zero, but very small anomalous
         form factors 
 $f_L-1, f_R, g_L, g_R \neq 0$, typically $\lesssim 0.01$. In particular
   the phases of these form factors due to final-state
   interactions or non-standard CP
  violation turn out to be small. As to $f_L$, a deviation 
 $\delta f_L \sim 0.1$ is possible if new, heavy 
 quarks with charge $Q=2/3$ exist that mix with the top quark
   \cite{delAguila:2000rc}. If a
4th sequential quark generation $t', b'$ exists then one expects
$|f_L| = |V_{tb}| <1$. A scan
  using input from $B,D,K$ decays and  electroweak precision
measurements yields  $|f_L| = |V_{tb}| >0.93$ 
  \cite{Eberhardt:2010bm}.
 A more exotic possibility is the existence
   of a  new heavy vector-like $T$ quark 
   as predicted, for instance, by  Little Higgs models or  by
 extra-dimension models. Mixing of $t$ and $T$ would reduce $f_L$;
 one expects  $|f_L| \gtrsim 0.9$. If a significant reduction of
  $f_L$  with respect to its SM value would be measured, it would point
  to the existence of a heavy $Q=2/3$ quark.

What about top decay modes other than $t\to bW$? In the SM all such
modes are rare; for instance the CKM-suppressed modes $t\to s,d$ have
  branching fractions  B($t\to W^+ s$) = $ 1.9 \times
10^{-3}$,  B($t\to W^+ d$) $ = 10^{-4}$. In SM extensions new decay
modes are possible, notably the decay into a light charged Higgs boson,
  $t\to bH^+$, and the decay into a light stop and a neutralino,  
$ t\to  {\tilde t}_1 {\tilde\chi}^0_1$. Searches for these modes
  at the Tevatron have been negative so far. At the LHC these modes
  will either be seen or excluded. Another issue
   are flavor-changing neutral currents
   involving the top quark, i.e. the existence of $t\to c,u$
   transitions. CDF has obtained the upper bound $B(t\to  Zq)<0.037$,
   while D$\emptyset$ has recently
    extracted the upper bounds $B(t\to gu)<2.0\times 10^{-4}$
        and ${\rm B}(t \to gc)<3.9 \times 10^{-3}$ from their
        single-top event sample \cite{Abazov:2010qk}. 
In the SM the branching fractions of
        the FCNC modes $t\to c,u$ are unmeasurably tiny due to almost
        perfect GIM cancellations; but even in many of the popular SM
        extensions  the branching fractions
     of these modes are, in view of the phenomenological constraints on
     these models, typically
      $\lesssim 10^{-5}$. (For a review, see e.g. \cite{Bernreuther:2008ju}.)
     If   branching ratios  
 ${\rm B}(t \to Z c) \gtrsim \times 10^{-4}$ would be found, it
 would point to mixing of $t$ with exotic (e.g. vector-like) quark(s).
 If a neutral Higgs boson $h$ lighter than the top quark with FCNC
 couplings exists then ${\rm B}(t \to h c) \sim  10^{-3}$ 
  and ${\rm B}(t \to gc) \sim  10^{-4}$ are possible.

\subsection{Charge asymmetry at the  Tevatron} 
The inclusive top (versus antitop) quark charge 
  asymmetry $A_t$ in $\ttbar$ production 
 at the Tevatron has recently stirred much interest, because the
  measurements do not quite match the SM expectations. In the
  SM this asymmetry is induced at NLO QCD (predominantly through 
$q{\bar q}\to \ttbar$). Assuming CP invariance, $A_t$ amounts to a 
  top-quark forward-backward asymmetry $A_{FB}^t$. 
   At NLO QCD $A_t = 0.051(6)$ \cite{Kuhn:1998kw,Antunano:2007da}, 
   while a related pair-asymmetry
      was computed to be  $A^{t {\bar t}}= 0.078(9)$ \cite{Antunano:2007da}.
          (Some weak-interaction
   contributions are included in these predictions.)
      For kinematical reasons  $A^{t {\bar t}}$ is larger than  $A_t$. 
  Resummation of QCD threshold logarithms \cite{Almeida:2008ug,Ahrens:2010zv}
 do not change these predictions
   significantly, but lead to a more realistic estimate of the
  theory uncertainty, namely  $\sim 15 - 20 \%$. It should be
  emphasized that all these predictions are made at the level of $\ttbar$
   states  without acceptance cuts.

At the Tevatron the top quark charge asymmetry
 was measured for $\ell$ + $j$ final states. D$\emptyset$
  obtained  $A^{\ttbar} =0.12 \pm 0.08 \pm 0.01$ \cite{:2007qb}. 
   This result has not
  been unfolded, while CDF has unfolded their data and obtained
  $A_{FB}^t=0.193\pm 0.065 \pm  0.024$ in their analysis of 2009
            \cite{CDFpublic1,datta}.
  Although there is no statistical significant discrepancy between
   experiments and the SM predictions, the present situation 
         leaves ample room for speculations about
      possible new physics contributions to $\ttbar$ production which
      (so far) show up only in this distribution. During the last two
      years many papers have appeared that address this issue, 
  see \cite{Rodrigo}.
 
Here, I want to add two remarks. A  top quark charge asymmetry
 induces an asymmetry $A^{\ell}$ for the charged leptons $\ell^\pm$ from
 $t$ and $\bar t$ decay  in
 dileptonic and in  $\ell$ + $j$ final states. Likewise, the $\ttbar$
 pair asymmetry leads to a leptonic pair asymmetry $A^{\ell\ell}$
   in the dileptonic
 sample. (See \cite{Bernreuther:2010ny} for the precise definitions  of
            $A^{\ell}$ and $A^{\ell\ell}$.)
  These leptonic asymmetries were computed to NLO QCD with
 respect to $\ttbar$ production and $t, \bar t$ decay, with weak
 interaction corrections and full NLO $\ttbar$ spin correlations
 included \cite{Bernreuther:2010ny}. 
   When standard Tevatron acceptance cuts are applied,
  these asymmetries are $A^{\ell}= 0.034(4)$ and 
 $A^{\ell\ell} = 0.044(4)$. (Only the
  uncertainties due to scale variations are given.)  
   These asymmetries are smaller than the corresponding asymmetries
   at the level of the intermediate $\ttbar$, because i) the 
  charged lepton does not
   follow the direction of its mother particle, ii)  the acceptance
   cuts diminish the asymmetries,  and iii) the $\ttbar$ spin correlations
    do have some effect on  $A^{\ell\ell}$. So far there are
   no experimental results on  $A^{\ell}$ and $A^{\ell\ell}$
   available. These asymmetries should be measurable more easily and
   with a higher precision than the above top-charge asymmetries.
   This may provide a more conclusive comparison with the SM
   results. \\
   Second, suppose there is a new physics contribution to the
    top-quark charge asymmetry. This new interaction need not be
     $C$- or $P$-violating in order to
            generate a non-zero contribution
 to $A_t$, but if it is $P$-violating, it would, in addition to
 contributing to $A_t$, also polarize the $t$ and $\bar t$ quarks
 of the $\ttbar$ sample in the production plane to some degree.
  (The $t$, $\bar t$   polarizations  due to the standard weak-interaction
   contributions to $\ttbar$ production
          is less than 1 $\%$.)  This can
    be checked by measuring the distributions
   $\sigma^{-1}d\sigma/ d\cos\theta_{\ell^\pm}$  in
    $\ell$ + $j$ final states, where $\theta_{\ell^\pm}$ are the
    ${\ell^\pm}$ helicity angles with respect to the
        top rest frame. The NLO SM predictions of these
      distributions are given in \cite{Bernreuther:2010ny}. Without acceptance cuts they would
      be essentially flat, but the acceptance cuts distort these distributions
    in the backward region $\cos\theta_{\ell}<0$.
       A sizeable  longitudinal polarization  of 
           the (anti)top-quark sample would result in  a non-flat distribution 
    in the forward region  $\cos\theta_{\ell}>0$. As the
    Tevatron experiments have already accumulated $\sim 10^3$ lepton +
     jets events, these $\ell^\pm$ distributions should be measurable
      with reasonable precision.

\subsection{$\ttbar$ spin correlations} 
  Top spin effects, in particular $\ttbar$ spin correlations
   are a rather unique feature of top quark physics, as compared to 
  the physics of lighter quarks. Final-state angular distributions and
  correlations induced by top-spin effects are ``good'' observables
   because this quark does not hadronize. Final state angular
   correlations induced bt $\ttbar$ spin correlations contain
   information about the 
   $\ttbar$ production and decay dynamics. Assuming
   that $t\to bW$ is the only decay mode of the top quark (possibly with small
   anomalous couplings) then a  closer look reveals that these
   correlations, especially the dilepton angular
    correlations,  essentially probe the $\ttbar$ production dynamics
   only. In the SM, the correlation of the $t$ and $\bar t$ spins is
   predominantly a QCD effect which is induced already at Born level.
    The degree of correlation depends, for a specific production
    reaction/dynamics, on the reference axes with respect to which
     the $t$ and $\bar t$ spin states are defined. At the Tevatron the
     SM-induced  $\ttbar$  spin correlations are largest in the
     so-called off-diagonal and beam bases, while at the LHC
      the helicity basis and an opening angle distribution defined in
      a specific way are good choices. These correlations were
      predicted to NLO QCD for dileptonic, $\ell$ + jets, and all jets
      final states \cite{Bernreuther:2001rq}. 
       These predictions were recently updated,
      taking weak-interaction contributions and acceptance cuts into 
   account \cite{Bernreuther:2010ny}. 
       The measurements of D$\emptyset$ and CDF   
   \cite{D0public1,CDFpublic2,CDFpublic3,Head}  agree
    with these predictions within the still large  experimental
    errors.

 At the LHC $\ttbar$ spin correlations should eventually be measurable
with significantly higher precision due to much larger data samples.
   If this will be the case then these observables will
   serve their purpose, namely to provide a further tool for exploring
   the $\ttbar$ production dynamics in detail. As mentioned, the angular
    correlations in the helicity basis and the opening angle
    distribution  ${\sigma}^{-1}{d\sigma/d\cos\varphi}$
      are good choices for detecting
     the SM-induced  $\ttbar$ spin correlations
       at  the LHC,  both for the $\ell\ell'$ 
     and     $\ell$ + jets final states.
                    As was  shown recently for the case
     of the LHC, also the   
 $\ell\ell'$  azimuthal angle correlation 
  ${\sigma}^{-1}{d\sigma}/{d\Delta\phi}$ (where
       $\Delta\phi = \phi^+ - \phi^-$) measured in the laboratory frame
 discriminates between correlated and uncorrelated $\ttbar$ events
 if only events with low pair-invariant mass $\mtt$ 
  are taken into account  \cite{Mahlon:2010gw,Mahlon}. 
  For the LHC at 14 TeV a useful cut is  
    $\mtt \leq 400$ GeV.  
     It was shown that the NLO QCD corrections
         to this distribution 
        in $\Delta\phi$ are sizeable, but its
         power to discriminate between correlated and uncorrelated 
 $\ttbar$ events remains \cite{Bernreuther:2010ny}.
 At the  LHC (14 TeV)   the ratio 
$\sigma_{\ell\ell'}(\mtt <400 \, {\rm GeV})/\sigma_{\ell\ell'} \simeq 18.6 \% $
  Thus with with an integrated luminosity of  1 ${\rm fb}^{-1}$
  one expects  $\sim$ 3200 dilepton events with low $\mtt$ before
  event selection. It remains to be investigated how much luminosity
  has to be accumulated in order to measure  this distribution at the
    level of several percent. 

Clearly, the $\Delta\phi$ distribution 
  is easier to measure than the opening angle
distribution or the helicity  correlation which require the
reconstruction of the $t$ and $\bar t$ rest frames. However, 
 the shapes of the $\Delta\phi$ distributions for correlated and
 uncorrelated events
           depend sensitively on how precisely ${\mtt}^{\rm cut}$ can
           be experimentally determined, and this distribution
           looses its 
  discriminating power rapidly for  ${\mtt}^{\rm cut} >$ 400 GeV.
  On the other hand the opening angle and helicity observables
  discriminate by design between correlated and 
 uncorrelated $\ttbar$ events, irrespective of whether or not 
  they are evaluated for all events or whether a maximum 
   ($<\mtt^{\rm cut}$)  or 
   minimum ($>\mtt^{\rm cut}$) selection cut is
 applied. While 
 ${\sigma}^{-1}{d\sigma}/{d\Delta\phi}$  
   probes the $\ttbar$ spin dynamics in the low-energy
      tail of the $\mtt$ spectrum, the
  helicity  and the opening angle correlation
       can be used  also
for the high energy tail, where  (non)resonant new
        physics effects may show up. The latter observables should 
   also be measured 
for $\ell +j$ events at the LHC. Although  the sensitivity decreases by
  a factor of about 2 for these channels, the data 
   samples will be about 6 times larger.

\subsection{Single top production} 
 The hadronic production of of single (anti)top quarks 
 is interesting for a number of reasons. These include: i) 
 The weak interactions are involved, and this provides a unique
    opportunity to directly
   explore  the top
   quark's charged current interactions. In the SM, the production cross
  section $\sigma_t \propto |V_{tb}|^2$. In fact, a closer look
  shows that all squared matrix elements $|V_{tq}|^2$
     of the 3rd row of the CKM matrix are involved.   ii) Due to the production
  mechanism, the single (anti)top samples are highly polarized. This
  remains to be exploited for the investigation of both the
  production and the decay dynamics.  iii) Single top production 
  is sensitive to BSM interactions that differ from those that can be
  traced in $\ttbar$ production. The single top productions modes may
   be affected by $t$- and $s$-channel exchanges of new, heavy charged
   resonances or by FCNC interactions. iv) By the time the production 
   cross sections will have been measured with sufficient precision
    and the partonic production mechanisms will have been pinned down,
     the data can be used for a direct determination of the $b$-quark
     content of the proton. 

In the SM the three main production reactions  are $t$-channel
 $W$-boson exchange (which is the dominant mode both at the Tevatron
and the LHC), $s$-channel $W$-boson  exchange, and the $tW$ production
mode. At the time of this workshop the 
        $t$ + $\bar t$ production cross
       sections  were measured by  D$\emptyset$ and CDF with a respective
   uncertainty of $\delta \sigma^{t+{\bar t}} \sim 25 \%$.
   The goal for the LHC (14 TeV), where the signal to background ratios
   become more favorable, is to measure the $t$-channel cross section
    with a precision of $\lesssim 10 \%$, which would amount
   to determining the strength $f_L$ of the $tWb$ vertex to $\sim 5\%$
   accuracy, provided the theoretical description
    is sufficiently precise. In view of the large backgrounds
    the present  and future analyses of single top events at
   the Tevatron and the LHC depend heavily on theory, i.e. on the
    calculated cross sections and distributions and their
    implementation in Monte Carlo codes. At present, in the 5-flavor
    scheme, the three $2\to 2$ single top production modes are known 
     to NLO CQD \cite{Stelzer:1997ns,Harris:2002md,Campbell:2004ch,Cao:2004ky,Cao:2005pq,White:2009yt} (plus threshold 
 resummations \cite{Kidonakis:2006bu}, plus 
  weak-interaction corrections \cite{Beccaria:2006ir,Beccaria:2008av}); the dominant $t$-channel production
   mode was computed to NLO QCD also in the 4-flavor scheme \cite{Campbell:2009gj} (where
     one considers $q g\to q't {\bar b}$ to be the leading order
     process).  The status of 
   the perturbative calculations and of 
      the NLO Monte Carlo codes will be discussed 
  by \cite{Rfrederix} and by \cite{PNason}, respectively. In spite of
      these impressive results there is still quite some work to do on the
      theory side in order to reach the goal of 
  exploring single top physics at the LHC at the level of $\sim 5\%$. 
 For the time being we are eager to learn about the expectations  
 of rediscovering single top events at the LHC \cite{Hischbuel}.

\subsection{New heavy resonances 
 $X_J \to \ttbar$ in ``early'' LHC phase?} 
 SM extensions  and/or alternatives to the SM Higgs mechanism, 
 e.g. supersymmetric extensions, 
top-conden\-sation and technicolor models, or models that involve extra
dimensions predict new  heavy resonances, some of which 
 couple (strongly) to top quarks. 
 Examples are  neutral or  charged  non-SM Higgs bosons, technicolor
   or top-color bound states, Kaluza-Klein (KK) excitations, heavy
  $t'$ and $b'$ quarks, or a heavy stop ${\tilde t}$. 
      At the Tevatron CDF and
  D$\emptyset$ have searched for such resonances and have  set
  mass/coupling  limits for instance on a heavy $W'$, $H^+$, $t'$,
  $b'$, and  ${\tilde t}$.

The $\ttbar$ invariant mass distribution
  is the key observable  in the
    search for electrically 
   neutral bosonic resonances $X_J$ that couple to $t \bar
  t$.  CDF and
  D$\emptyset$ have not found a significant excess in the measured
  $\mtt$ spectrum up to $\sim$ 1 TeV compared to the SM expectation.
  The searches  for $p {\bar p} \to X_J \to t \bar t$
 led to the exclusion of a leptophobic $Z'$ boson (which appears in TC2
 models) with  mass $M_{Z'}< 820$ GeV and of massive KK gluons
   with mass $M_G \lesssim 1$ TeV, see \cite{sinervo}.
 It will take a while before these limits/sensitivity ranges will by
  superseded by the LHC experiments. 

Most of the above-mentioned SM extensions contain
  heavy, neutral Higgs bosons or 
   Higgs-like spin-zero resonances $\phi$ in their physical particle
  spectrum.  For instance 2-Higgs doublet extensions or the MSSM
  predict 3 neutral Higgs bosons, and both models allow for the
   possibility that two of the three states have a mass of about
    $2 m_t$ or larger.  In the case of a pseudoscalar state
    $\phi = A$ ($J^{PC}= 0^{-+}$) there is the specific feature
    that  $A$ does not couple to   $W^+ W^- , ZZ$ in lowest order
    because of $CP$ mismatch. In addition, the loop-induced
     decays into the golden channels $A\to W^+ W^- , ZZ$ are
      suppressed in many models in large portions 
         of their parameter spaces \cite{Bernreuther:2010uw}. 
     But $A$ can strongly couple to
      top quarks, like the other states $\phi$. The most likely
      production mode of a $\phi$ resonance is gluon fusion. 
             The amplitude of $gg\to \phi \to \ttbar$
      interferes with  the amplitude of the non-resonant $gg \to
      \ttbar$ background, which  leads to a typical peak-dip
resonance structure in the $M_{t\bar t}$ spectrum
        \cite{Dicus:1994bm,Bernreuther:1997gs}.
If  such a state $\phi$ exists, with a mass in the
   range 300 GeV $\lesssim m_\varphi \lesssim {\cal
  O}(600 \, {\rm GeV})$ and with a strong Yukawa coupling to the
   top quark, then it is conceivable that it would be seen
   as a resonance bump at the LHC, but not at the Tevatron!
   It remains to be investigated how precisely
  the $\mtt$ spectrum can be measured  after the first
 LHC (7 TeV) running period.

\section{Outlook}
The top quark, the heaviest known fundamental particle, offers the
  unique possibility to explore the interactions  of a bare quark
   at distances below the attometer scale. The future of top quark physics
   is certainly bright. As far as top quark physics at the LHC in its present
    operating mode   is
   concerned we have to face reality. We are all happy that the LHC is
   running at 7 TeV, and we have to see how much integrated
    luminosity will have been delivered to the experiments 
         by the end of 2011; perhaps 
     $\sim 200$ pb$^{-1}$  or up to   $\sim 1$ fb$^{-1}$ ?  
Needless to say: the ATLAS and CMS experiments 
 first have to calibrate their detectors and software
  tools with the recorded data.
 What kind of top physics can we expect? We will learn about
   it in the talks \cite{HYoo,Lessard,Ferrari}. If one stays on the
      pessimistic side assuming only    
  $L \sim 200$ pb$^{-1}$ by the end of 2011, then 
  from  the $\ttbar$ cross section 
$\sigma_{\ttbar}\simeq 150$ pb  at 7 TeV one would  have
30 k $\ttbar$ events before selection, i.e. roughly 
  $\sim 200 \ell \ell'$ and $\sim$ 2k $\ell\, + \,  j$ events after
 selection. When will the first single Euro tops be observed? 
 Is this possible within the present  LHC running period?
The single-top cross section 
   $\sigma_t \simeq$  65 pb implies  $\geq 13$ k  tops before selection
if $L\geq  200$ pb$^{-1}$.
 What will come more from CDF and D0?
These questions will hopefully be addressed, too,  in the next days. 
 We all look forward to a week of stimulating talks and discussions.

\subsection{Acknowledgments}
This work was  supported by Deutsche Forschungsgemeinschaft, SFB TR9.

\end{document}